\documentclass[sigconf]{acmart}

\AtBeginDocument{%
  \providecommand\BibTeX{{%
    \normalfont B\kern-0.5em{\scshape i\kern-0.25em b}\kern-0.8em\TeX}}}

\usepackage{amsmath}
\DeclareMathOperator*{\argmax}{arg\,max}

\usepackage{subfigure}

\begin{document}

\settopmatter{printacmref=false} 
\renewcommand\footnotetextcopyrightpermission[1]{} 
\pagestyle{plain} 

\title{EnDSUM: Entropy and Diversity based Disaster Tweet Summarization}

\author{Piyush Kumar Garg}
\affiliation{%
  \institution{IIT Patna}
  \country{India}
}
\email{piyush_2021cs05@iitp.ac.in}
\author{Roshni Chakraborty}
\affiliation{%
  \institution{Aalborg University}
  \country{Denmark}
}
\email{roshnic@cs.aau.dk}

\author{Sourav Kumar Dandapat}
\affiliation{%
  \institution{IIT Patna}
  \country{India}
}
\email{sourav@iitp.ac.in}

\begin{abstract}
 The huge amount of information shared in Twitter during disaster events are utilized by government agencies and humanitarian organizations to ensure quick crisis response and provide situational updates. However, the huge number of tweets posted makes manual identification of the relevant tweets impossible. To address the information overload, there is a need to automatically generate summary of all the tweets which can highlight the important aspects of the disaster. In this paper, we propose an entropy and diversity based summarizer, termed as \textit{EnDSUM}, specifically for disaster tweet summarization. Our comprehensive analysis on $6$ datasets indicates the effectiveness of \textit{EnDSUM} and additionally, highlights the scope of improvement of \textit{EnDSUM}.
\end{abstract}


\maketitle

\section{Introduction} \label{s:intro}

\par Social media platforms, like Twitter, are highly important mediums of information during disasters. For example, humanitarian organizations and government agencies rely on Twitter to identify relevant information on different categories, such as affected population, urgent need of resources, infrastructure damage, etc~\cite{imran2015towards}. However, the huge number of tweets posted and the high vocabulary diversity~\cite{castillo2016big,chakraborty2018predicting} make it a challenging to manually find the relevant information~\cite{vieweg2014integrating, imran2015processing}. In order to address this issue, several research works~\cite{rudra2015extracting,dutta2018ensemble} have proposed specific tweet summarization approaches for disaster events.  


\par Existing disaster tweet summarization approaches could be segregated into content based~\cite{rudra2015extracting}, graph based~\cite{dutta2018ensemble}, deep learning based~\cite{dusart2021tssubert}, and ontology based~\cite{garg2022ontorealsumm} approaches on the basis of the mechanism they follow. While content based approaches~\cite{rudra2015extracting, rudra2018extracting} rely on only the importance of the words present in a tweet to determine its selection to the summary, deep learning based approaches~\cite{dusart2021tssubert} consider both content and contextual importance of the tweet. However, none of these approaches consider the vocabulary diversity and therefore, fails to always ensure diversity in summary and coverage of all the important categories present in the tweets. In order to address these, graph based approaches~\cite{dutta2018ensemble, dutta2015graph} initially group similar tweets together such that each group represents a category by community detection algorithms, thereby handling the vocabulary diversity followed by selecting representative tweets from each group to create the summary to ensure coverage. However, automatic community detection algorithms fails to automatically segregate the tweets into different categories due to the vocabulary overlap among tweets of different categories. Therefore, Garg et al. ~\cite{garg2022ontorealsumm} initially identify the category of each tweet by an ontology based approach and then, select tweets from each category to generate the summary. However, none of these approaches try to handle the vocabulary diversity simultaneously while selecting the tweets into the summary. For example, these existing approaches are dependent on identifying the categories initially which lead to bad summaries, such as reduced diversity in summary, if the categories are not identified correctly.

\par In order to resolve this, we propose \textit{EnDSUM}, an entropy and diversity based disaster summarizer where we automatically select that tweet into summary which provides the best information coverage of all the tweets, i.e., \textit{entropy} and most novel information, i.e., \textit{diversity}. Therefore, \textit{EnDSUM} can generate the summary automatically without explicitly identifying the category of a tweet. Although there are few single and multiple document summarization approaches~\cite{khurana2022investigating, feigenblat2017unsupervised, aji2012document, luo2010effectively, ravindra2004multi} that have highlighted the relevance of \textit{entropy} based selection of sentences into summary, those approaches are not directly applicable to disaster tweets. The reason being the informal structure of tweets, absence of storyline in tweets and the high vocabulary diversity in user generated tweets. Our evaluation of \textit{EnDSUM} with existing state-of-the-art disaster tweet summarization approaches on $6$ different disasters shows its high effectiveness on $5$ datasets. However, we observe that the performance of \textit{EnDSUM} degrades when there is considerable vocabulary overlap among the tweets which belong to different categories of the same disaster event. The reason being we consider only content based information for calculation of \textit{entropy} and \textit{diversity}. The organization of the paper is as follows. We discuss problem definition and proposed approach in Section~\ref{s:selection} followed by the experiment details in Section~\ref{s:exp} and conclusions in Section~\ref{s:con}.

\begin{table*}[ht]
    \centering 
    \caption{F1-score of ROUGE-1, ROUGE-2 and ROUGE-L score of \textit{EnDSUM} and baselines on $6$ datasets is shown.}
    \label{table:Result}
    \resizebox{\textwidth}{!}{\begin{tabular}{|c|c|c|c|c|c|c|c|c|c|} \hline
    
        \textbf{Dataset} & \textbf{Approaches} & \textbf{ROUGE-1} & \textbf{ROUGE-2} & \textbf{ROUGE-L} & \textbf{Dataset} & \textbf{Approaches} & \textbf{ROUGE-1} & \textbf{ROUGE-2} & \textbf{ROUGE-L}  \\ \cline{3-5} \cline{8-10}
        
         &  & \textbf{F1-score} & \textbf{F1-score} & \textbf{F1-score} &  &  & \textbf{F1-score} & \textbf{F1-score} & \textbf{F1-score}  \\ \hline
        
                 & $EnDSUM$ & 0.55 & 0.21 & 0.27 &                       & $EnDSUM$ & {\bf0.51} & {\bf0.16} & {\bf0.24}\\\cline{2-5}\cline{7-10}
                 & $B_1$    & 0.49 & 0.22 & 0.29 &                       & $B_1$    & 0.20 & 0.04 & 0.20 \\ \cline{2-5}\cline{7-10}
        ${D_1}$  & $B_2$    & 0.48 & 0.18 & 0.25 &            ${D_4}$    & $B_2$    & 0.47 & 0.14 & 0.21 \\ \cline{2-5}\cline{7-10}
                 & $B_3$    & 0.52 & 0.21 & 0.23 &                       & $B_3$    & 0.45 & 0.11 & 0.21 \\ \cline{2-5}\cline{7-10}
                 & $B_4$    & {\bf0.56} & {\bf0.23} & {\bf0.29} &        & $B_4$    & 0.50 & 0.15 & 0.23 \\ \hline
        
                 & $EnDSUM$ & {\bf0.52} & {\bf0.17} & {\bf0.24} &        & $EnDSUM$ & {\bf0.52} & {\bf0.13} & {\bf0.24} \\ \cline{2-5}\cline{7-10}
                 & $B_1$    & 0.48 & 0.13 & 0.22 &                       & $B_1$    & 0.19 & 0.04 & 0.18 \\ \cline{2-5}\cline{7-10}
        ${D_2}$  & $B_2$    & 0.47 & 0.14 & 0.22 &            ${D_5}$    & $B_2$    & 0.48 & 0.10 & 0.20 \\ \cline{2-5}\cline{7-10}
                 & $B_3$    & 0.44 & 0.12 & 0.22 &                       & $B_3$    & 0.50 & 0.12 & 0.22 \\ \cline{2-5}\cline{7-10}
                 & $B_4$    & 0.49 & 0.15 & 0.23 &                       & $B_4$    & 0.51 & 0.13 & 0.22 \\ \hline

                 & $EnDSUM$ & {\bf0.52} & 0.14 & {\bf0.26} &            & $EnDSUM$ & {\bf0.55} & {\bf0.27} & {\bf0.44} \\ \cline{2-5}\cline{7-10}
                 & $B_1$    & 0.45 & 0.13 & 0.23 &                      & $B_1$    & 0.53 & 0.26 & 0.33 \\ \cline{2-5}\cline{7-10}
        ${D_3}$  & $B_2$    & 0.46 & 0.14 & 0.24 &            ${D_6}$   & $B_2$    & 0.52 & 0.22 & 0.29 \\ \cline{2-5}\cline{7-10}
                 & $B_3$    & 0.44 & 0.14 & 0.23 &                      & $B_3$    & 0.48 & 0.20 & 0.27 \\ \cline{2-5}\cline{7-10}
                 & $B_4$    & 0.48 & {\bf0.16} & 0.25 &                 & $B_4$    & 0.51 & 0.20 & 0.29 \\ \hline
    \end{tabular} }
\end{table*}

\section{Proposed Approach} \label{s:selection}

\par Given a disaster event, $E$, that consists of $m$ tweets, $T$ = \{$T_1, T_2,..., T_m\}$, we aim to prepare a summary, $S$, by selecting $L$ tweets from $T$ such that it provides the maximum information coverage from $T$ with minimum redundant information in the final summary. Therefore, we propose \textit{EnDSUM} where we iteratively selecting the tweet that can ensure the maximum entropy of all the tweets and maximum diversity in summary. While selection of the tweet with maximum \textit{entropy} ensures information coverage of a category, selection of the tweet with the maximum \textit{diversity} ensures not multiple tweets from the same category are selected \cite{chakraborty2019tweet,chakraborty2017network}. Therefore, at every iteration, we select the tweet ($T^*$), which has the maximum score by Equation~\ref{eq:objf}. 

\begin{align}
        T^* = \operatorname{\argmax} \sum_{T_i \in T}\alpha * E(T_i) + \beta * D(T_i, S^{'})  
        \label{eq:objf}
\end{align}

where, $E(T_i)$ represents the \textit{entropy} of tweet, $T_i$, and $D(T_i, S^{'})$ represents the information \textit{diversity} provided by $T_i$ with respect to the already selected tweets in summary, $S^{'}$. $\alpha$ and $\beta$ are the tunable parameters which represent the importance of $E(T_i)$ and $D(T_i, S^{'})$ respectively. We consider $\alpha$ and $\beta$ as $0.5$ to provide equal importance to both \textit{entropy} and \textit{diversity}. Although there are several available mechanisms to calculate $E(T_i)$, we rely on \textit{Karci Entropy} ~\cite{hark2020karci} for \textit{EnDSUM}. \textit{Karci Entropy} can resolve the inherent vocabulary diversity in disaster tweets as it calculates the \textit{entropy} of a tweet, $E(T_i)$, by considering the similarity of $T_i$ with the other tweets as shown in Equation \ref{eq:entropy}.

\begin{align}
    E(T_i,K) = \sum_{j=1}^{\lvert K \rvert } \vert -P_{ij}^\gamma \log P_{ij} \vert, \ 0<\gamma
    \label{eq:entropy}
\end{align}

where, $\gamma$ represents the importance of similarity. We consider $\gamma$ as $0.5$ as highlighted by Hark et al.~\cite{hark2020karci}. $K$ is the list of similar tweets of $T_i$, where a tweet is said to be similar to $T_i$ if the content based cosine similarity, i.e. $P_{ij}$ between them is higher than $0$ (as shown in \cite{hark2020karci}) and $P_{ij}$ is the normalized number of overlapping between $T_i$ and $T_j$ normalized by the total number of overlapping keywords of $T_i$ with any tweet. We calculate $D(T_i, S^{'})$ as (1-$Sim(T_i, S^{'})$) where $Sim(T_i, S^{'})$ represents the overlap in keywords between $T_i$ and $S^{'}$ by 

\begin{align}
    Sim(T_i, S^{'}) =  \sum_{k \in S^{'}} \frac{Overlap(T_i, T_k)}{Length(T_i)} 
    \label{eq:gain}
\end{align}

where,  $Length(T_i)$ is the number of keywords of $T_i$. We follow Khan et al.~\cite{khan2013multi} to identify the keywords of $T_i$ as the nouns, verbs, adjectives present in $T_i$ and similarly, for $S^{'}$, we consider the distinct set of nouns, verbs, adjectives present in all the tweets of $S^{'}$. Therefore, a lower $Sim(T_i, S^{'})$ ensures $T_i$ has minimum redundant content information with respect to already generated summary, $S^{'}$, and a higher $E(T_i)$ ensures $T_i$ has higher information coverage of the category. 

\section{Experiments and Results}\label{s:exp}
\par In this Section, we provide details of the experiment and results. For the datasets, we consider \textit{Los Angeles International Airport Shooting}~\footnote{https://en.wikipedia.org/wiki/2013\_Los\_Angeles\_International\_Airport\_shooting} (\textit{$D_1$}) provided by Olteanu et al.~\cite{olteanu2015expect}, \textit{Hurricane Matthew}~\footnote{https://en.wikipedia.org/wiki/Hurricane\_Matthew} (\textit{$D_2$}), \textit{Puebla Mexico Earthquake}~\footnote{https://en.wikipedia.org/wiki/2017\_Puebla\_earthquake} (\textit{$D_3$}), \textit{Pakistan Earthquake}~\footnote{https://en.wikipedia.org/wiki/2019\_Kashmir\_earthquake} (\textit{$D_4$}) and \textit{Midwestern U.S. Floods}~\footnote{https://en.wikipedia.org/wiki/2019\_Midwestern\_U.S.\_floods} (\textit{$D_5$}) provided by Alam et al.~\cite{Alam2021humaid} and \textit{Sandy Hook Elementary School Shooting}~\footnote{https://en.wikipedia.org/wiki/Sandy\_Hook\_Elementary\_School\_shooting} (\textit{$D_6$}) provided by Dutta et al.~\cite{dutta2018ensemble}. We perform lemmatization, convert to lower case and remove of Twitter specific keywords~\cite{arachie2020unsupervised} and retweets as pre-processing. We consider the ground truth summary provided by Garg et al.~\cite{garg2022ontorealsumm} for $D_1$-$D_5$ and by Dutta et al.~\cite{dutta2018ensemble} for $D_6$. 
We compare \textit{EnDSUM} with content based ~\cite{rudra2019summarizing} (\textit{$B_1$}), graph based~\cite{dutta2018ensemble} (\textit{$B_2$}), sub-event based~\cite{rudra2018identifying} (\textit{$B_3$}) and ontology based~\cite{garg2022ontorealsumm} (\textit{$B_4$}) disaster summarization approaches.
    
\textit{Results and Discussion} : We evaluate the performance of \textit{EnDSUM} and the existing research with the ground truth summary using ROUGE-N~\cite{lin2004rouge} F1-score score when N=$1$, $2$, and L. Our observations from Table~\ref{table:Result} indicate that \textit{EnDSUM} ensures better ROUGE-N F1-score over all baselines for $D_2$-$D_6$. The improvement is highest over $B_1$ baseline and lowest over $B_4$ baseline. \textit{EnDSUM} performs worse than $B_4$ for Rouge-N scores and worse than $B_1$ for Rouge-2 and Rouge-L scores on $D_1$. Therefore, although \textit{EnDSUM} has highly effective performance in most scenarios, it sometimes fails to resolve the vocabulary overlap across different categories in a disaster, as seen for $D_1$. Therefore, to resolve this, we are working towards making  \textit{EnDSUM} resilient irrespective of the vocabulary diversity by considering semantic and contextual similarity along with the already considered content similarity for entropy and diversity calculation.

\section{Conclusions and Future Works}\label{s:con}

\par In this paper, we propose a novel entropy and diversity based tweet summarizer, \textit{EnDSUM} for disaster events. Our experimental analysis on $6$ disaster datasets indicates both the effectiveness of \textit{EnDSUM} and its scope of improvement. For example, to handle the the high vocabulary overlap among categories, we are working to both include semantic and contextual similarity while calculating entropy and diversity in \textit{EnDSUM}. Furthermore, while most summarization algorithms generate a predefined length summary, we intend to extend \textit{EnDSUM} such that it provides complete information coverage of the disaster event. For example, intuitively, the summary length varies on the basis of the information diversity in a disaster event, therefore a summary of length less than the required length leads to less information coverage whereas a summary with more number of tweets than the required will reduce the information diversity. We believe by incorporating these changes in \textit{EnDSUM}, it would provide an effective performance irrespective of the disaster.

\bibliographystyle{ACM-Reference-Format}
\bibliography{sample-base}


\begin{thebibliography}{26}


\ifx \showCODEN    \undefined \def \showCODEN     #1{\unskip}     \fi
\ifx \showDOI      \undefined \def \showDOI       #1{#1}\fi
\ifx \showISBNx    \undefined \def \showISBNx     #1{\unskip}     \fi
\ifx \showISBNxiii \undefined \def \showISBNxiii  #1{\unskip}     \fi
\ifx \showISSN     \undefined \def \showISSN      #1{\unskip}     \fi
\ifx \showLCCN     \undefined \def \showLCCN      #1{\unskip}     \fi
\ifx \shownote     \undefined \def \shownote      #1{#1}          \fi
\ifx \showarticletitle \undefined \def \showarticletitle #1{#1}   \fi
\ifx \showURL      \undefined \def \showURL       {\relax}        \fi
\providecommand\bibfield[2]{#2}
\providecommand\bibinfo[2]{#2}
\providecommand\natexlab[1]{#1}
\providecommand\showeprint[2][]{arXiv:#2}

\bibitem[Aji and Kaimal(2012)]%
        {aji2012document}
\bibfield{author}{\bibinfo{person}{Subhanpurno Aji} {and}
  \bibinfo{person}{Ramachandra Kaimal}.} \bibinfo{year}{2012}\natexlab{}.
\newblock \showarticletitle{Document summarization using positive pointwise
  mutual information}.
\newblock \bibinfo{journal}{\emph{AIRCC's International Journal of Computer
  Science and Information Technology}} \bibinfo{volume}{4}, \bibinfo{number}{2}
  (\bibinfo{year}{2012}), \bibinfo{pages}{47--55}.
\newblock


\bibitem[Alam et~al\mbox{.}(2021)]%
        {Alam2021humaid}
\bibfield{author}{\bibinfo{person}{Firoj Alam}, \bibinfo{person}{Umair Qazi},
  \bibinfo{person}{Muhammad Imran}, {and} \bibinfo{person}{Ferda Ofli}.}
  \bibinfo{year}{2021}\natexlab{}.
\newblock \showarticletitle{HumAID: Human-Annotated Disaster Incidents Data
  from Twitter with Deep Learning Benchmarks}.
\newblock \bibinfo{journal}{\emph{arXiv preprint arXiv:2104.03090}}
  (\bibinfo{year}{2021}).
\newblock


\bibitem[Arachie et~al\mbox{.}(2020)]%
        {arachie2020unsupervised}
\bibfield{author}{\bibinfo{person}{Chidubem Arachie}, \bibinfo{person}{Manas
  Gaur}, \bibinfo{person}{Sam Anzaroot}, \bibinfo{person}{William Groves},
  \bibinfo{person}{Ke Zhang}, {and} \bibinfo{person}{Alejandro Jaimes}.}
  \bibinfo{year}{2020}\natexlab{}.
\newblock \showarticletitle{Unsupervised detection of sub-events in large scale
  disasters}. In \bibinfo{booktitle}{\emph{Proceedings of the AAAI Conference
  on Artificial Intelligence}}, Vol.~\bibinfo{volume}{34}.
  \bibinfo{pages}{354--361}.
\newblock


\bibitem[Castillo(2016)]%
        {castillo2016big}
\bibfield{author}{\bibinfo{person}{Carlos Castillo}.}
  \bibinfo{year}{2016}\natexlab{}.
\newblock \bibinfo{booktitle}{\emph{Big crisis data: social media in disasters
  and time-critical situations}}.
\newblock \bibinfo{publisher}{Cambridge University Press}.
\newblock


\bibitem[Chakraborty et~al\mbox{.}(2017)]%
        {chakraborty2017network}
\bibfield{author}{\bibinfo{person}{Roshni Chakraborty}, \bibinfo{person}{Maitry
  Bhavsar}, \bibinfo{person}{Sourav Dandapat}, {and} \bibinfo{person}{Joydeep
  Chandra}.} \bibinfo{year}{2017}\natexlab{}.
\newblock \showarticletitle{A network based stratification approach for
  summarizing relevant comment tweets of news articles}. In
  \bibinfo{booktitle}{\emph{International Conference on Web Information Systems
  Engineering}}. Springer, \bibinfo{pages}{33--48}.
\newblock


\bibitem[Chakraborty et~al\mbox{.}(2019)]%
        {chakraborty2019tweet}
\bibfield{author}{\bibinfo{person}{Roshni Chakraborty}, \bibinfo{person}{Maitry
  Bhavsar}, \bibinfo{person}{Sourav~Kumar Dandapat}, {and}
  \bibinfo{person}{Joydeep Chandra}.} \bibinfo{year}{2019}\natexlab{}.
\newblock \showarticletitle{Tweet summarization of news articles: An objective
  ordering-based perspective}.
\newblock \bibinfo{journal}{\emph{IEEE Transactions on Computational Social
  Systems}} \bibinfo{volume}{6}, \bibinfo{number}{4} (\bibinfo{year}{2019}),
  \bibinfo{pages}{761--777}.
\newblock


\bibitem[Chakraborty et~al\mbox{.}(2018)]%
        {chakraborty2018predicting}
\bibfield{author}{\bibinfo{person}{Roshni Chakraborty},
  \bibinfo{person}{Abhijeet Kharat}, \bibinfo{person}{Apalak Khatua},
  \bibinfo{person}{Sourav~Kumar Dandapat}, {and} \bibinfo{person}{Joydeep
  Chandra}.} \bibinfo{year}{2018}\natexlab{}.
\newblock \showarticletitle{Predicting Tomorrow's Headline using Twitter
  Deliberations.}. In \bibinfo{booktitle}{\emph{CIKM Workshops}}.
\newblock


\bibitem[Dusart et~al\mbox{.}(2021)]%
        {dusart2021tssubert}
\bibfield{author}{\bibinfo{person}{Alexis Dusart}, \bibinfo{person}{Karen
  Pinel-Sauvagnat}, {and} \bibinfo{person}{Gilles Hubert}.}
  \bibinfo{year}{2021}\natexlab{}.
\newblock \showarticletitle{TSSuBERT: Tweet Stream Summarization Using BERT}.
\newblock \bibinfo{journal}{\emph{arXiv preprint arXiv:2106.08770}}
  (\bibinfo{year}{2021}).
\newblock


\bibitem[Dutta et~al\mbox{.}(2018)]%
        {dutta2018ensemble}
\bibfield{author}{\bibinfo{person}{Soumi Dutta}, \bibinfo{person}{Vibhash
  Chandra}, \bibinfo{person}{Kanav Mehra}, \bibinfo{person}{Asit~Kumar Das},
  \bibinfo{person}{Tanmoy Chakraborty}, {and} \bibinfo{person}{Saptarshi
  Ghosh}.} \bibinfo{year}{2018}\natexlab{}.
\newblock \showarticletitle{Ensemble algorithms for microblog summarization}.
\newblock \bibinfo{journal}{\emph{IEEE Intelligent Systems}}
  \bibinfo{volume}{33}, \bibinfo{number}{3} (\bibinfo{year}{2018}),
  \bibinfo{pages}{4--14}.
\newblock


\bibitem[Dutta et~al\mbox{.}(2015)]%
        {dutta2015graph}
\bibfield{author}{\bibinfo{person}{Soumi Dutta}, \bibinfo{person}{Sujata
  Ghatak}, \bibinfo{person}{Moumita Roy}, \bibinfo{person}{Saptarshi Ghosh},
  {and} \bibinfo{person}{Asit~Kumar Das}.} \bibinfo{year}{2015}\natexlab{}.
\newblock \showarticletitle{A graph based clustering technique for tweet
  summarization}. In \bibinfo{booktitle}{\emph{2015 4th international
  conference on reliability, infocom technologies and optimization
  (ICRITO)(trends and future directions)}}. IEEE, \bibinfo{pages}{1--6}.
\newblock


\bibitem[Feigenblat et~al\mbox{.}(2017)]%
        {feigenblat2017unsupervised}
\bibfield{author}{\bibinfo{person}{Guy Feigenblat}, \bibinfo{person}{Haggai
  Roitman}, \bibinfo{person}{Odellia Boni}, {and} \bibinfo{person}{David
  Konopnicki}.} \bibinfo{year}{2017}\natexlab{}.
\newblock \showarticletitle{Unsupervised query-focused multi-document
  summarization using the cross entropy method}. In
  \bibinfo{booktitle}{\emph{Proceedings of the 40th International ACM SIGIR
  Conference on research and development in information retrieval}}.
  \bibinfo{pages}{961--964}.
\newblock


\bibitem[Garg et~al\mbox{.}(2022)]%
        {garg2022ontorealsumm}
\bibfield{author}{\bibinfo{person}{Piyush~Kumar Garg}, \bibinfo{person}{Roshni
  Chakraborty}, {and} \bibinfo{person}{Sourav~Kumar Dandapat}.}
  \bibinfo{year}{2022}\natexlab{}.
\newblock \showarticletitle{OntoRealSumm: Ontology based Real-Time Tweet
  Summarization}.
\newblock \bibinfo{journal}{\emph{arXiv preprint arXiv:2201.06545}}
  (\bibinfo{year}{2022}).
\newblock


\bibitem[Hark and Karc{\i}(2020)]%
        {hark2020karci}
\bibfield{author}{\bibinfo{person}{Cengiz Hark} {and} \bibinfo{person}{Ali
  Karc{\i}}.} \bibinfo{year}{2020}\natexlab{}.
\newblock \showarticletitle{Karc{\i} summarization: A simple and effective
  approach for automatic text summarization using Karc{\i} entropy}.
\newblock \bibinfo{journal}{\emph{Information Processing \& Management}}
  \bibinfo{volume}{57}, \bibinfo{number}{3} (\bibinfo{year}{2020}),
  \bibinfo{pages}{102187}.
\newblock


\bibitem[Imran and Castillo(2015)]%
        {imran2015towards}
\bibfield{author}{\bibinfo{person}{Muhammad Imran} {and}
  \bibinfo{person}{Carlos Castillo}.} \bibinfo{year}{2015}\natexlab{}.
\newblock \showarticletitle{Towards a data-driven approach to identify
  crisis-related topics in social media streams}. In
  \bibinfo{booktitle}{\emph{Proceedings of the 24th International Conference on
  World Wide Web}}. \bibinfo{pages}{1205--1210}.
\newblock


\bibitem[Imran et~al\mbox{.}(2015)]%
        {imran2015processing}
\bibfield{author}{\bibinfo{person}{Muhammad Imran}, \bibinfo{person}{Carlos
  Castillo}, \bibinfo{person}{Fernando Diaz}, {and} \bibinfo{person}{Sarah
  Vieweg}.} \bibinfo{year}{2015}\natexlab{}.
\newblock \showarticletitle{Processing social media messages in mass emergency:
  A survey}.
\newblock \bibinfo{journal}{\emph{ACM Computing Surveys (CSUR)}}
  \bibinfo{volume}{47}, \bibinfo{number}{4} (\bibinfo{year}{2015}),
  \bibinfo{pages}{1--38}.
\newblock


\bibitem[Khan et~al\mbox{.}(2013)]%
        {khan2013multi}
\bibfield{author}{\bibinfo{person}{Muhammad Asif~Hossain Khan},
  \bibinfo{person}{Danushka Bollegala}, \bibinfo{person}{Guangwen Liu}, {and}
  \bibinfo{person}{Kaoru Sezaki}.} \bibinfo{year}{2013}\natexlab{}.
\newblock \showarticletitle{Multi-tweet summarization of real-time events}. In
  \bibinfo{booktitle}{\emph{2013 International Conference on Social
  Computing}}. \bibinfo{publisher}{IEEE}, \bibinfo{pages}{128--133}.
\newblock


\bibitem[Khurana and Bhatnagar(2022)]%
        {khurana2022investigating}
\bibfield{author}{\bibinfo{person}{Alka Khurana} {and} \bibinfo{person}{Vasudha
  Bhatnagar}.} \bibinfo{year}{2022}\natexlab{}.
\newblock \showarticletitle{Investigating Entropy for Extractive Document
  Summarization}.
\newblock \bibinfo{journal}{\emph{Expert Systems with Applications}}
  \bibinfo{volume}{187} (\bibinfo{year}{2022}), \bibinfo{pages}{115820}.
\newblock


\bibitem[Lin(2004)]%
        {lin2004rouge}
\bibfield{author}{\bibinfo{person}{Chin-Yew Lin}.}
  \bibinfo{year}{2004}\natexlab{}.
\newblock \showarticletitle{Rouge: A package for automatic evaluation of
  summaries}. In \bibinfo{booktitle}{\emph{Text summarization branches out}}.
  \bibinfo{pages}{74--81}.
\newblock


\bibitem[Luo et~al\mbox{.}(2010)]%
        {luo2010effectively}
\bibfield{author}{\bibinfo{person}{Wenjuan Luo}, \bibinfo{person}{Fuzhen
  Zhuang}, \bibinfo{person}{Qing He}, {and} \bibinfo{person}{Zhongzhi Shi}.}
  \bibinfo{year}{2010}\natexlab{}.
\newblock \showarticletitle{Effectively leveraging entropy and relevance for
  summarization}. In \bibinfo{booktitle}{\emph{Asia Information Retrieval
  Symposium}}. Springer, \bibinfo{pages}{241--250}.
\newblock


\bibitem[Olteanu et~al\mbox{.}(2015)]%
        {olteanu2015expect}
\bibfield{author}{\bibinfo{person}{Alexandra Olteanu}, \bibinfo{person}{Sarah
  Vieweg}, {and} \bibinfo{person}{Carlos Castillo}.}
  \bibinfo{year}{2015}\natexlab{}.
\newblock \showarticletitle{What to expect when the unexpected happens: Social
  media communications across crises}. In \bibinfo{booktitle}{\emph{Proceedings
  of the 18th ACM conference on computer supported cooperative work \& social
  computing}}. \bibinfo{pages}{994--1009}.
\newblock


\bibitem[Ravindra et~al\mbox{.}(2004)]%
        {ravindra2004multi}
\bibfield{author}{\bibinfo{person}{G Ravindra}, \bibinfo{person}{N
  Balakrishnan}, {and} \bibinfo{person}{KR Ramakrishnan}.}
  \bibinfo{year}{2004}\natexlab{}.
\newblock \showarticletitle{Multi-document automatic text summarization using
  entropy estimates}. In \bibinfo{booktitle}{\emph{International Conference on
  Current Trends in Theory and Practice of Computer Science}}. Springer,
  \bibinfo{pages}{289--300}.
\newblock


\bibitem[Rudra et~al\mbox{.}(2018a)]%
        {rudra2018extracting}
\bibfield{author}{\bibinfo{person}{Koustav Rudra}, \bibinfo{person}{Niloy
  Ganguly}, \bibinfo{person}{Pawan Goyal}, {and} \bibinfo{person}{Saptarshi
  Ghosh}.} \bibinfo{year}{2018}\natexlab{a}.
\newblock \showarticletitle{Extracting and summarizing situational information
  from the twitter social media during disasters}.
\newblock \bibinfo{journal}{\emph{ACM Transactions on the Web (TWEB)}}
  \bibinfo{volume}{12}, \bibinfo{number}{3} (\bibinfo{year}{2018}),
  \bibinfo{pages}{1--35}.
\newblock


\bibitem[Rudra et~al\mbox{.}(2015)]%
        {rudra2015extracting}
\bibfield{author}{\bibinfo{person}{Koustav Rudra}, \bibinfo{person}{Subham
  Ghosh}, \bibinfo{person}{Niloy Ganguly}, \bibinfo{person}{Pawan Goyal}, {and}
  \bibinfo{person}{Saptarshi Ghosh}.} \bibinfo{year}{2015}\natexlab{}.
\newblock \showarticletitle{Extracting situational information from microblogs
  during disaster events: a classification-summarization approach}. In
  \bibinfo{booktitle}{\emph{Proceedings of the 24th ACM International on
  Conference on Information and Knowledge Management}}.
  \bibinfo{pages}{583--592}.
\newblock


\bibitem[Rudra et~al\mbox{.}(2019)]%
        {rudra2019summarizing}
\bibfield{author}{\bibinfo{person}{Koustav Rudra}, \bibinfo{person}{Pawan
  Goyal}, \bibinfo{person}{Niloy Ganguly}, \bibinfo{person}{Muhammad Imran},
  {and} \bibinfo{person}{Prasenjit Mitra}.} \bibinfo{year}{2019}\natexlab{}.
\newblock \showarticletitle{Summarizing situational tweets in crisis scenarios:
  An extractive-abstractive approach}.
\newblock \bibinfo{journal}{\emph{IEEE Transactions on Computational Social
  Systems}} \bibinfo{volume}{6}, \bibinfo{number}{5} (\bibinfo{year}{2019}),
  \bibinfo{pages}{981--993}.
\newblock


\bibitem[Rudra et~al\mbox{.}(2018b)]%
        {rudra2018identifying}
\bibfield{author}{\bibinfo{person}{Koustav Rudra}, \bibinfo{person}{Pawan
  Goyal}, \bibinfo{person}{Niloy Ganguly}, \bibinfo{person}{Prasenjit Mitra},
  {and} \bibinfo{person}{Muhammad Imran}.} \bibinfo{year}{2018}\natexlab{b}.
\newblock \showarticletitle{Identifying sub-events and summarizing
  disaster-related information from microblogs}. In
  \bibinfo{booktitle}{\emph{The 41st International ACM SIGIR Conference on
  Research \& Development in Information Retrieval}}.
  \bibinfo{pages}{265--274}.
\newblock


\bibitem[Vieweg et~al\mbox{.}(2014)]%
        {vieweg2014integrating}
\bibfield{author}{\bibinfo{person}{Sarah Vieweg}, \bibinfo{person}{Carlos
  Castillo}, {and} \bibinfo{person}{Muhammad Imran}.}
  \bibinfo{year}{2014}\natexlab{}.
\newblock \showarticletitle{Integrating social media communications into the
  rapid assessment of sudden onset disasters}. In
  \bibinfo{booktitle}{\emph{International Conference on Social Informatics}}.
  Springer, \bibinfo{pages}{444--461}.
\newblock


\end{thebibliography}
\end{document}